\documentclass[12pt] {article}
\pdfoutput=1

\usepackage[hyphens,spaces,obeyspaces]{url}
\usepackage{graphicx}
\usepackage{amsmath}
\usepackage{caption}
\usepackage{tabularx}
\usepackage{amssymb}
\usepackage{blindtext}
\usepackage{enumerate}
\usepackage{hyperref}
\usepackage{mathtools}
\usepackage{marvosym}
\usepackage{caption,subcaption}
\usepackage{multirow}
\usepackage{datetime}
\usepackage{acronym}
\usepackage{balance}
\usepackage{fancybox}
\usepackage{colortbl}
\usepackage{array}
\usepackage{mystyle}
\usepackage{setspace}
\usepackage{cite}
\usepackage{epstopdf}
\usepackage{epsfig}
\usepackage{amssymb}
\usepackage{multirow}
\usepackage{algorithm}
\usepackage{algpseudocode}
\usepackage{lipsum}
\usepackage{acronym}
\usepackage{threeparttable}
\usepackage{longtable}
\usepackage{listings, lstautogobble}
\usepackage{booktabs}
\usepackage{threeparttable}
\usepackage[table,xcdraw]{xcolor}
\usepackage{wrapfig}
\usepackage{paralist}
\usepackage{enumitem}
\usepackage[utf8]{inputenc}
\usepackage[english]{babel}
\usepackage[nottoc]{tocbibind}
\usepackage[obeyspaces,hyphens,spaces]{url}
\usepackage[english]{babel}
\providecommand{\keywords}[1]{\textbf{\textit{Keywords---}} #1}

{\begin{list}{}%
         {\setlength{\leftmargin}{#1}}%
         \item[]%
}
{\end{list}}

\newdateformat{monthyeardate}{%
  \monthname[\THEMONTH], \THEYEAR}

\begin{document}
\sloppy

\title{\Large{\textbf{A Formally Verified HOL4 Algebra for Event Trees} }}
\author{Mohamed Abdelghany, Waqar Ahmad, and Sofi\`ene Tahar\vspace*{2em}\\
Department of Electrical and Computer Engineering,\\
Concordia University, Montr\'eal, QC, Canada 
\vspace*{1em}\\
\{m\_eldes,waqar,tahar\}@ece.concordia.ca
 \vspace*{3em}\\
\textbf{TECHNICAL REPORT}\\
\date{May 2020}
}
\maketitle

\newpage
\begin{abstract}

Event Tree (ET) analysis is widely used as a forward deductive safety analysis technique for decision-making at the critical-system design stage. ET is a schematic diagram representing all possible operating states and external events in a system so that one of these possible scenarios can occur. In this report, we propose to use the HOL4 theorem prover for the formal modeling and step-analysis of ET diagrams. To this end, we developed a formalization of ETs in higher-order logic, which is based on a generic \textit{list}-datatype that can: (i) construct an arbitrary level of ET diagrams; (ii) reduce the irrelevant ET branches; (iii)~partition ET paths; and (iv)~perform the probabilistic analysis based on the occurrence of certain events. For illustration purposes, we conduct the formal ET stepwise analysis of an electrical power grid and also determine its System Average Interruption Frequency Index~($\mathcal{SAIFI}$), which is an important indicator for system~reliability. \\ 
\end{abstract}

\pagebreak

\section{Introduction}
\label{Sec:Intro}

Nowadays, the fulfillment of stringent safety requirements for critical-systems, which are prevalent, e.g., in smart grids and automotive industry, has been encouraging safety design engineers to use formal techniques as per recommendations of safety standards, such as IEC 61850 \cite{mackiewicz2006overview} and ISO 26262 \cite{palin2011iso}. Therefore, it is required to build necessary formal support for rigorous reliability analysis so that they become an essential step in the design process and ensure the delivery of a trusted service without failures \cite{bozzano2010design}. Several reliability modeling techniques have been developed, such as Fault Trees (FT) \cite{towhidnejad2002fault}, Reliability Block Diagrams (RBD) \cite{brall2007reliability} and Event Trees~(ET) \cite{papazoglou1998mathematical}, that describe the behavior of components for a given system. FTs mainly provide a graphical model for analyzing the factors causing a system failure upon their occurrences. On the other hand, RBDs allow us to model the success relationships of complex systems. ETs enumerate all possible operating states and external events in a system in the form of a tree structure represented by an \textit{initiating node} and \textit{branches}~\cite{papazoglou1998mathematical}. The results of the ET analysis are extremely useful for safety analysts as it provide a more detailed system view compared to FTs and RBDs.\\

Papazoglou \cite{papazoglou1998mathematical} was the first researcher to lay down the mathematical foundations of ET in the late 90s. He described the ET analysis in \textit{four} main steps: (1)~\textit{Generation}: construct a complete ET model; (2) \textit{Reduction}: removal of unnecessary ET branches; (3)~\textit{Partitioning}: extract a collection of ET paths according to the system failure and success events; and lastly (4)~\textit{Probabilistic analysis}:~ evaluate the probabilities of ET paths based on the occurrence of a certain event. But the analysis of ET proposed in \cite{papazoglou1998mathematical} is done purely manually using a paper-and-pencil approach. On the other hand, there exist several commercial tools based on Monte-Carlo Simulation for ET analysis, such as ITEM~\cite{ITEM_tp}, Isograph~\cite{Isograph_tp}, and EC Tree~\cite{ECTREE_tp}, which have been widely used to determine sequentially failure and success scenarios of real-world systems, like electrical power grids \cite{muzik2018possibilities}, nuclear power plants~\cite{peplow2004calculating} and railways \cite{ku2011reliability}. Prior to utilizing these tools for ET analysis, the users must draw a given system actual ET diagram manually, maybe on paper. Both of these approaches may introduce inaccuracies in the ET analysis due to human infallibility and analysis approximations caused by the numerical methods in the simulation tools, respectively. A more efficient and practical way is to define functions describing the pattern of modeling ETs as well as ET probabilistic properties.  \\

In this report, we propose to use HOL theorem proving \cite{HOL_tp}, which provides us the ability to accurately model and also rigorously verify the essential ET properties. For this purpose, we endeavor to formalize the four steps of ET analysis using the HOL4 proof assistant, i.e., generation, reduction, partitioning and probabilistic analysis, as described by Papazoglou \cite{papazoglou1998mathematical}. We present two syntactically different, but semantically equivalent formalizations for ET analysis, using \textit{set} and \textit{list}-datatypes, respectively. The former \textit{set}-datatype ET formalization is described by Papazoglou, however, it cannot mimic the graphical model of an ET consisting of an initiating node and branches since the elements in sets are orderless. The ordering is important in \textit{Steps} 3 and 4 (\textit{reduction} and \textit{partitioning} processes) of the ET analysis. In the latter approach, the \textit{list}-datatype inherently preserves the index of its member elements and naturally captures the graphical structure of ETs. Also, from our experience, the reasoning about ET reduction and partitioning properties using the \textit{set}-datatype is quite cumbersome and significantly slow compared to the \textit{list}-datatype especially when the ET diagram becomes tremendously large. Therefore, we use the \textit{list}-datatype to formalize all \textit{four} steps of ET analysis in HOL4. For that purpose, we propose to use the \textit{list}-datatype that inherently preserves the index of its member elements and naturally captures the graphical structure of ETs. For illustration purposes, we conduct the formal ET analysis of a practical power grid system consisting of $\mathcal{N}$ transmission lines and $\mathcal{M}$~customers. Subsequently, we also formally determine the System Average Interruption Frequency Index~($\mathcal{SAIFI}$)~\cite{grainger2003power}, which  is an important reliability index describing the average frequency of interruptions in an electrical power~systems.  \\

The rest of the report is organized as follows: In Section 2, we present the related literature review. Section 3 briefly summarize the fundamentals of ETs. In Section~4, we present the details of our HOL4 formalization of ETs using the \textit{set}-datatype. Section~5 describes the formalization of ETs by developing a new recursive datatype \small{\texttt{EVENT\_TREE}}. In Section 7, we present the formalization of ETs reduction and partitioning. Section 6 describes the formal probabilistic analysis of ETs. In Section 8, we present the formal ET analysis-based of a power grid system and the assessment of its reliability index $\mathcal{SAIFI}$. Lastly, Section~9 concludes the report.

\vspace{-3mm}
\section{Related Work}
\label{Sec: Related Work}

Only a few work have previously considered using formal methods to model and analyze ETs. For instance, N{\`y}vlt~et~al. in~\cite{nyvlt2012dependencies} used Petri nets to model the cascading failure of sub-systems and their effect on the entire system using the standard FT and ET modeling techniques. The authors proposed a new method based on P-invariants to obtain a model of cascading dependencies in ETs \cite{nyvlt2012dependencies}. However, according to the authors, they are not able to obtain verified equations from that model \cite{nyvlt2012dependencies}. HOL4 \cite{HOL_tp} has been previously used by Ahmad et al. in \cite{ahmad2015} to formalize FTs and RBDs. The FT formalization includes a new datatype consisting of \texttt{AND}, \texttt{OR} and \texttt{NOT} FT constructors \cite{towhidnejad2002fault} to analyze the factors causing a system failure. Similarly, Ahmad et al. in~\cite{ahmed2016formalization} defined a new \texttt{RBD} datatype to model and analyze the success relationships of a system using different RBD configurations~\cite{brall2007reliability}, such as series, parallel and combination of series and parallel. However, both of these formalizations are limited to analyzing either a system failure or its success only. On the other hand, ETs have the superior ability to analyze both failure and success scenarios in a system. For the formalization of ET in HOL4, the existing \texttt{treeTheory} in the standard library of HOL4 only allows drawing a specific tree with leaves and nodes manually. To the best of our knowledge, this is the first work, which develops a formal modeling and step-analysis of ETs using~HOL4 theorem prover.\\

\section{Event Trees}
\label{Sec: event tree analysis}

\begin{figure}[!t]
	\includegraphics[scale= 0.35]{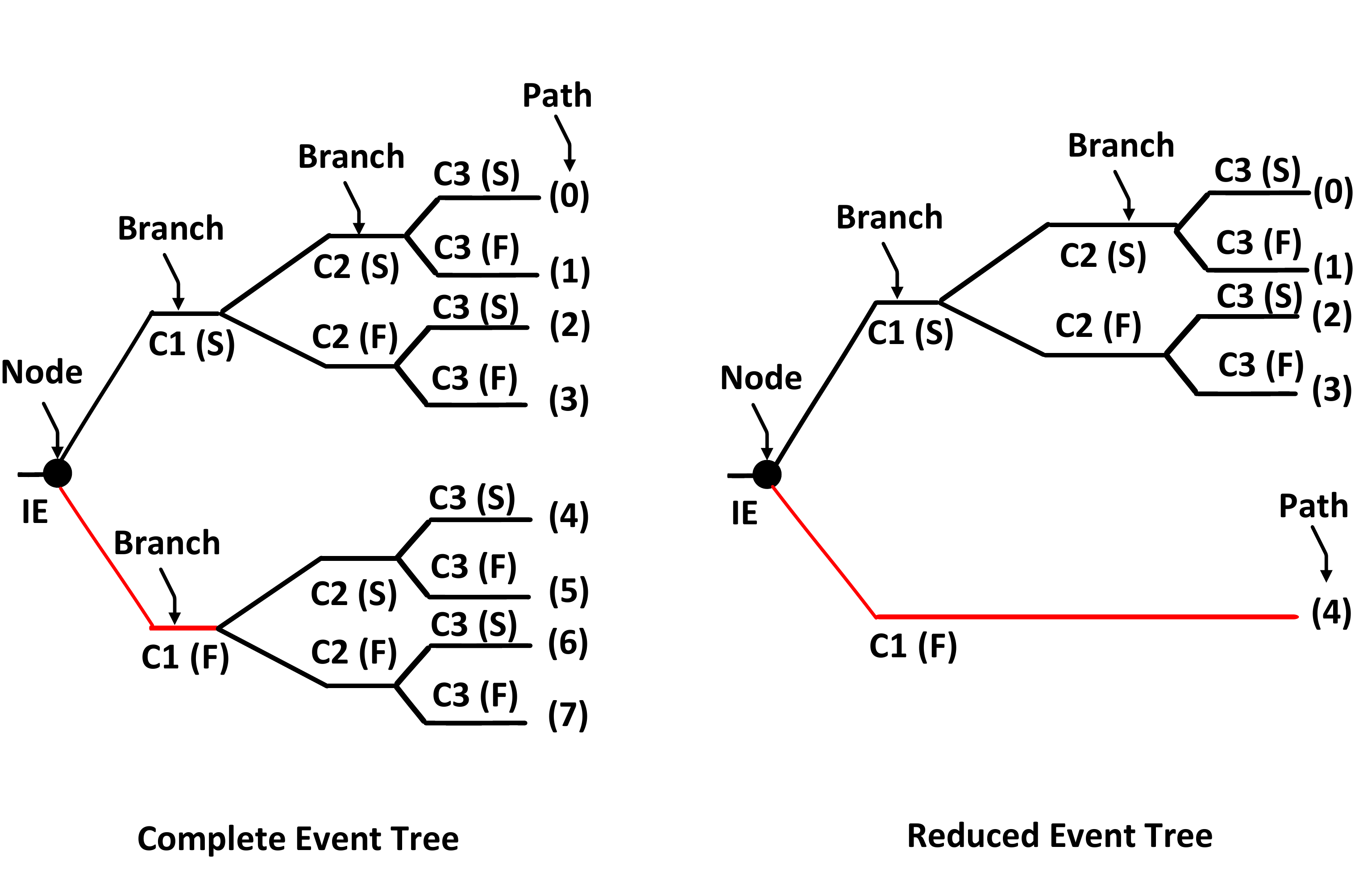} 
	\centering
	\caption{A sample ET diagrams / S (Success state) / F (Failure state)}
	\label{eventtree}
\end{figure}

An ET diagram is a graphical model that enumerate all possible combinations of component states and external events in a system in the form of a tree structure. ETs utilize the forward logic \cite{hu1999evaluating} starting by an Initiating Event (IE) called~\textit{node} and then all possible scenarios of an event are drawn as \textit{branches}. For instance, consider a system consist of three components $C_1$, $C_2$ and $C_3$, each has two operational states, i.e., operating or failing. The ET \textit{four} step-analysis defined by Papazoglou \cite{papazoglou1998mathematical} are as follows:
\begin{enumerate}

\item \textit{Generation}: Construct a complete ET diagram that draws all possible scenarios, which is well-known as \textit{paths}.  Each \textit{path} consists of a unique sequence of events. Fig.~\ref{eventtree} depicts 8 paths (0-7) with all possible scenarios that can occur.
\item  \textit{Reduction}: Model the accurate functional behavior of the system in the sense that the irrelevant branches should be removed from a complete ET. This can be done by deleting some specific branches corresponding to the occurrence of certain events, which are known as \textit{Complete Cylinders}~(CCs) \cite{papazoglou1998functional}. These cylinders are ET \textit{paths} consisting of $\mathcal{N}$~events and they are conditional on the occurrence of $\mathcal{K}$~\textit{Conditional Events}~(CEs) in their respective paths and they are referred to as CCs  with respect to $\mathcal{K}$~\cite{papazoglou1998functional}.  For instance, if the critical-component $C_1$ fails then the whole system fails regardless of the status of the rest of the components, i.e., $C_2$ and $C_3$, as shown in Fig. \ref{eventtree}. Therefore, paths 4-7 are CCs with respect to~$C_{1F}$.
\item  \textit{Partitioning}: partition of an ET diagram is essential as we are only interested in the occurrence of certain events according to the system failure and success events. For instance, suppose we are only focusing on the failure of the system  in Fig. \ref{eventtree}, then ET paths 3 and 4 are obtained from the reduced ET.

\item  \textit{Probabilistic analysis}: Lastly, evaluate the probabilities of ET paths based on the occurrence of a certain event. These probabilities represent the likelihood of each scenario that is possible to occur in a system so that \textit{only one} can occur~\cite{papazoglou1998mathematical}. This implies that all paths connected to a node are disjoint (mutually exclusive) \cite{papazoglou1998mathematical}. Assuming that all events in an ET are mutually independent that the probability of any ET path can be computed by simply multiplying the individual probabilities of all the events associated with it~\cite{papazoglou1998mathematical}. For example, the probability of the system failure in Fig. \ref{eventtree}, i.e., paths 3 and 4, is expressed mathematically as: 
\begin{equation}
	\centering 
	\mathcal{P} (System_{Failure}) = \mathcal{P} (C_{1S}) \times \mathcal{P} (C_{2F}) \times  \mathcal{P} (C_{3F}) + \mathcal{P} (C_{1F}) 
\end{equation}

\noindent where $\mathcal{P} (C_{XF})$ is the unreliability function or the probability of failure for a component \textit{X} and $\mathcal{P} (C_{XS})$ is the reliability function or the probability of operating.\\
\end{enumerate}

In the next sections, we present, in detail, the formalization of ETs using the \textit{set} and the \textit{list} data-types, respectively. The reason for using the \textit{set} theory is that most of the mathematical foundations of ETs from the work of Papazoglou~\cite{papazoglou1998mathematical} are built on sets. However, ordering of events in ET paths is important during \textit{Steps 2} and 3 of the ET analysis. Therefore, a sequence-preserving formalization of ETs in the \textit{list} theory should be adopted. In order to ensure the correspondence of the \textit{set} and \textit{list} theory based ET formalizations, we formally verify the equivalence between them.


\section{ET Formalization using Sets}
\label{Sub:ET Set Theory}

An event outcome space ($\mathcal{W}$) is referred to a set of all possible scenarios of an IE or modes of operation of a system critical-component, which must satisfy the following constraints according to Papazoglou \cite{papazoglou1998mathematical}:
\begin{enumerate}[label=\alph*)]
	\item \textit{Distinct}: All outcomes in an event outcome space must be unique.
	\item \textit{Disjoint (mutually exclusive)}: Any pair of events from a set of events outcome space cannot occur at the same~time.
	\item \textit{Complete}: An event outcome space must contain all possible events that can~occur. 
	\item \textit{Finite}: An event outcome space must consist of a finite number of elements.
	\begin{equation}
	\label{Eq:Omega}
	\centering 
	\mathcal{W}= \{\omega_j\} \qquad j = 1,2,\dots,\mathcal{N}
	\end{equation}
\end{enumerate}

\noindent We formalize the above-mentioned event outcome space ($\mathcal{W}$)  constraints in HOL4 as~follows:
\begin{flushleft}
	\vspace{1pt} \small{\texttt{\bf{type\_abbrev}}
		\texttt{(``event'', ``:($\alpha$ -> bool)'')}}
\end{flushleft}
\begin{flushleft}
\label{Definition 1}
	\texttt{\bf{Definition 1:}}\\
	\label{DEF: Event Space}
	\vspace{1pt} \small{\texttt{$\vdash$ $\Omega$ $\mathcal{W}$ =  \{x | x $\in$ $\mathcal{W}$ $\wedge$ 
			$\varnothing$ $\in$ $\mathcal{W}$ $\wedge$ 
			disjoint $\mathcal{W}$ $\wedge$  
			FINITE $\mathcal{W}$\}}}
\end{flushleft}
\noindent where $\mathcal{W}$ is a set of \textit{events} representing the possibilities resulting from an IE or modes of operation of a system component in HOL4. The elements in a set are intrinsically distinct and thus ensuring the constraint~(a). The function \texttt{disjoint} ensures that each pair of elements in a given set is mutually exclusive satisfying constraint~(b). The completeness of an event outcome space, constraint~(c), means containing all possible events or modes of a system component that can occur. In many practical systems, some components are redundant for improving system reliability and they are only used when required, i.e.,~in a hold state meaning neither success nor fail. This completeness of the event outcome space can be ensured by adding an empty set~$\varnothing$ representing the default (not in-use) case, i.e.,~a component is neither in success nor in failure state. The HOL4 function \texttt{FINITE} guarantees that the set of event outcome space must consist of a finite number of elements, as indicated by constraint~(d). \\

Consider a system having two events, say $\textit{E}_1$ and $\textit{E}_2$, with two event outcome spaces $\mathcal{W}_1$ and $\mathcal{W}_2$, respectively. The Cartesian product ($\bigotimes$) of these event outcome spaces returns a set of ($\mathcal{N}_1 \times \mathcal{N}_2$) pairs containing all possible outcome pairs for the occurrence of $\textit{E}_1$ and $\textit{E}_2$ together (i.e., $\mathcal{W}_1 \bigotimes \mathcal{W}_2$). In ET, an intersection operation is performed on each member of these pairs to obtain a valid event outcome space. In other words, the resulting event outcome space from the Cartesian product of two event outcome spaces also satisfies the above-mentioned constraints. We formalize this concept in HOL4 as follows:\\

\noindent We first construct a set by taking each element from the event outcome spaces $\mathcal{W}_1$ and $\mathcal{W}_2$ and then performing an intersection operation on these elements as:

\begin{flushleft}
\label{Definition 2}
	\texttt{\bf{Definition 2:}}\\
	\label{DEF: Inter Cartesian}
	\vspace{1pt} \small{\texttt{$\vdash$ $\mathcal{W}_1$ $\bigcap^\bigotimes$  $\mathcal{W}_2$ =  \{x $\cap$ y | x $\in$ $\Omega$ $\mathcal{W}_1$  $\wedge$ y $\in$ $\Omega$ $\mathcal{W}_2$\}}}
\end{flushleft}

\noindent Next, we ensure that the obtained duets from Definition 2 are mutually exclusive. For instance, consider two arbitrary outcomes ($\omega_{1m}$ $\cap$ $\omega_{2n}$) and ($\omega_{1k}$ $\cap$ $\omega_{2l}$) at least ($\textit{m}$ $\neq$ $\textit{k}$) or ($\textit{n}$ $\neq$ $\textit{l}$) must be true.

\begin{flushleft}
\label{Definition 3}
	\texttt{\bf{Definition 3:}}\\
	\label{DEF: Cartesian Product}
	\vspace{1pt} \small{\texttt{$\vdash$ $\mathcal{W}_1$ $\bigotimes$ $\mathcal{W}_2$ =  \{x | x $\in$ $\mathcal{W}_1$ $\bigcap^\bigotimes$  $\mathcal{W}_2$  $\wedge$ disjoint ($\mathcal{W}_1$ $\bigcap^\bigotimes$ $\mathcal{W}_2$)\}}}
\end{flushleft}

\noindent To ensure the validity of an event outcome space, as described in Eq. 2, we define a \textit{predicate} function in HOL4 as follows: 
\begin{flushleft}
\label{Definition 4}
	\texttt{\bf{Definition 4:}}\\
	\label{DEF: valid event space outcome}
	\vspace{1pt} \small{\texttt{$\vdash$ $\Omega^\vDash$ $\mathcal{W}$ $\Leftrightarrow$
				$\varnothing$ $\in$ $\mathcal{W}$ $\wedge$ disjoint $\mathcal{W}$ $\wedge$  
				FINITE $\mathcal{W}$}}
\end{flushleft}

Using the above definitions, we formally verify that the function $\bigotimes$ forms a valid event outcome space.

\begin{flushleft}
	\texttt{\bf{Theorem 1:}} (Cartesian product fulfilling the event outcome space constraints)\\
	\label{THM: CARTESIAN_PRODUCT_ITSELF}
	\vspace{1pt} \small{\texttt{$\vdash$ $\Omega^\vDash$ $\mathcal{W}_1$} $\wedge$
			   \texttt{$\Omega^\vDash$ $\mathcal{W}_2$} $\Rightarrow$ \texttt{$\Omega^\vDash$ ($\mathcal{W}_1$ $\bigotimes$ $\mathcal{W}_2$)}}
\end{flushleft}

Now, we can define a generic function as defined by Papazoglou~\cite{papazoglou1998mathematical} that can take an arbitrary set of event outcome spaces and generate the corresponding ET diagram (i.e., $\mathcal{W}_1 \bigotimes \mathcal{W}_2 \bigotimes \dots \bigotimes \mathcal{W}_\mathcal{N}$). For this purpose, we use the HOL4 function \texttt{ITSET} that can recursively apply $\bigotimes$  on a given set of event outcome spaces as~follows:

\begin{flushleft}
\label{Definition 5}
	\texttt{\bf{Definition 5:}}\\
	\label{DEF: N Cartesian Product}
	\vspace{1pt} \small{\texttt{$\vdash$ $\mathcal{S}$ $\bigotimes^\mathcal{N}$ $\mathcal{W}_\mathcal{N}$ = 
			 ITSET 
			\vspace{1pt} \small{\texttt{($\lambda$$\mathcal{W}_1$ $\mathcal{W}_2$. 
					$\mathcal{W}_1$  $\bigotimes$ $\mathcal{W}_2$)}}
			$\mathcal{S}$ $\mathcal{W}_\mathcal{N}$}}
\end{flushleft}

\noindent where $\mathcal{S}$ is a \textit{set} containing all event outcome spaces till $\mathcal{N}{-1}$ (i.e., $\mathcal{S}$ = \{$\mathcal{W}_1$; $\mathcal{W}_2$; \dots; $\mathcal{W}_{\mathcal{N}{-1}}$\}) and $\mathcal{W}_\mathcal{N}$ represents the last event outcome space. In order to reason about essential properties of above-mentioned ET model, we formally verify the following properties, by utilizing the properties of the HOL4 function \texttt{ITSET}, on a given set of event outcome spaces~as:

\begin{flushleft}
	\texttt{\bf{Theorem 2:}}\\
	\vspace{1pt} \small{\texttt{$\vdash$}
		\texttt{($\mathcal{W}_1$ INSERT $\mathcal{S}$) $\bigotimes^\mathcal{N}$ $\mathcal{W}_\mathcal{N}$ = ($\mathcal{S}$ DELETE $\mathcal{W}_1$) $\bigotimes^\mathcal{N}$ ($\mathcal{W}_1$  $\bigotimes$ $\mathcal{W}_\mathcal{N}$)}}		
\end{flushleft}

\begin{flushleft}
	\texttt{\bf{Theorem 3:}}\\
	\vspace{1pt} \small{\texttt{$\vdash$}
		\texttt{($\mathcal{W}_1$ INSERT $\mathcal{S}$) $\bigotimes^\mathcal{N}$ $\mathcal{W}_\mathcal{N}$ = $\mathcal{W}_1$ $\bigotimes$ (($\mathcal{S}$ DELETE $\mathcal{W}_1$) $\bigotimes^\mathcal{N}$ $\mathcal{W}_\mathcal{N}$)}}		
\end{flushleft}

The order of events in a path is irrelevant when evaluating the probabilities of a given path \cite{papazoglou1998functional}, i.e., the probability of path {($C_{1S}$, $C_{2F}$, $C_{3F}$)} in Eq. 1 is exactly equivalent to the probability of path {($C_{3F}$, $C_{2F}$, $C_{1S}$)} due to the commutative property of intersection and the events independence. However, it is important to preserve the order of events in ET paths during \textit{Steps} 2 and 3  (\textit{reduction} and \textit{partitioning}) of the ET analysis~\cite{papazoglou1998mathematical} while the elements in sets are orderless.  A possible way to resolve the problem of ordering in the \textit{set}-datatype is by assigning a unique number to each set element representing a branch during the ET modeling. However, when the ET diagram becomes tremendously large, the set-based reasoning is quite cumbersome and significantly slow compared to the \textit{list}-datatype. For that purpose, in the next three sections, we present the formalization of all \textit{four} ET analysis steps using the \textit{list} datatype, which inherently preserves the order of elements.

\vspace{-3mm}
\section{ET Formalization using Lists}
\label{Sub:ET List Theory}

We start the formalization of ETs by developing a new recursive datatype  \texttt{EVENT\_TREE} in HOL4 as follows:

\begin{flushleft}
	\vspace{1pt} \small{ \texttt{\bf{Hol\_datatype}}  
		\texttt{EVENT\_TREE =  \quad \hspace{-3mm} ATOMIC of ($\alpha$ event) \\
			\qquad \qquad \qquad \qquad \qquad \qquad \quad \quad \hspace{-3.5mm}| NODE of  EVENT\_TREE list \\ 
			\qquad \qquad \qquad \qquad \qquad \qquad \quad \quad \hspace{-3.5mm}| BRANCH of ($\alpha$ event) $\Rightarrow$ EVENT\_TREE list}}
\end{flushleft}

\noindent The type constructors \texttt{NODE} and \texttt{BRANCH} are recursive functions on \texttt{EVENT\_TREE}-typed lists. A semantic function is then defined over the \texttt{EVENT\_TREE} datatype that can yield a corresponding ET diagram as:
\begin{flushleft}
\label{Definition 6}
	\texttt{\bf{Definition 6:}}\\
	\vspace{1pt} 
	\small{\texttt{$\vdash$ ETREE (ATOMIC X) = X $\wedge$}} \\ 
	\small{\texttt{\quad ETREE (NODE (h::t)) = ETREE h $\cup$ (ETREE (NODE t))}}  $\wedge$ \\    	
	\small{\texttt{\quad ETREE (BRANCH X (h::t)) = X $\cap$ (ETREE h $\cup$ ETREE (BRANCH X t))}}    	
\end{flushleft}

\noindent The function \texttt{ETREE} takes a set \texttt{X}, identified by a type constructor \texttt{ATOMIC} and returns the given set \texttt{X}. If the function \texttt{ETREE} takes a list of type \texttt{EVENT\_TREE}, identified by a type constructor \texttt{NODE}, then it returns the union of all elements after applying the function \texttt{ETREE} on each element of the given list. Similarly, if the function \texttt{ETREE} takes a set \texttt{X} and a list of type \texttt{EVENT\_TREE}, identified by a type constructor \texttt{BRANCH}, then it performs the intersection of the set \texttt{X}  with the union of the head of the given list after applying the function \texttt{ETREE} and the recursive call of the \texttt{BRANCH} constructor.\\

To formally define a function that can model a complete ET for $\mathcal{N}$ lists, similar to Definition 5, we start by defining a function that can model an ET for two lists, say $\mathrm{L}_1$  and $\mathrm{L}_2$, in HOL4 as: 

\begin{flushleft}
\label{Definition 7}
	\texttt{\bf{Definition 7:}}\\
	\vspace{1pt} 
	\small{\texttt{$\vdash$ (h::t) $\bigotimes_{\mathrm{L}}$ $\mathrm{L}_2$ = BRANCH h $\mathrm{L}_2$::t $\bigotimes_{\mathrm{L}}$ $\mathrm{L}_2$}}   
	
\end{flushleft}
\noindent where the function $\bigotimes_{\mathrm{L}}$ takes two different \texttt{EVENT\_TREE}-typed lists and returns an \texttt{EVENT\_TREE}-typed list by recursively applying the \texttt{BRANCH} constructor on each element of the first list paired with the entire second list. \\

Now, we can define a generic function that takes an arbitrary list of event outcome spaces and generates a corresponding complete ET diagram, i.e., \textit{Step 1 (Generation)} of ET analysis~\cite{papazoglou1998mathematical}. For this purpose, we utilize the HOL4 function \texttt{FOLDR} that recursively applies $\bigotimes_{\mathrm{L}}$ on a given list of event outcome spaces as:
\begin{flushleft}
\label{Definition 8}
	\texttt{\bf{Definition 8:}}\\
	\vspace{1pt} 
	\small{\texttt{$\vdash$ L $\bigotimes^\mathcal{N}_{\mathrm{L}}$ $\mathrm{L}_\mathcal{N}$  = FOLDR (\small{\texttt{$\lambda$$\mathrm{L}_1$ $\mathrm{L}_2$. $\mathrm{L}_1$  $\bigotimes_{\mathrm{L}}$ $\mathrm{L}_2$}}) $\mathrm{L}_\mathcal{N}$ L}}
\end{flushleft}

\noindent where \textit{$\mathrm{L}$} is a \textit{list} of all event outcome spaces till $\mathcal{N}{-1}$ (i.e., L = [[$\mathcal{W}_1$]; [$\mathcal{W}_2$];\dots; [$\mathcal{W}_{\mathcal{N}{-1}}$]]) and $\mathrm{L}_\mathcal{N}$ = [$\mathcal{W}_\mathcal{N}$].\\

In order to ensure the correspondence of the \textit{list} and \textit{set} theory based ET formalizations, we formally verify the equivalence between Definitions 3 and 7 and Definitions~5 and 8, in HOL4 as:
\begin{flushleft}
	\texttt{\bf{Theorem 4:}}\\
	\vspace{1pt} 
	\small{\texttt{$\vdash$ $\Omega^\vDash_{\mathrm{L}}$ [$\mathrm{L}_1$;$\mathrm{L}_2$] \hspace{-1mm} $\Rightarrow$ ETREE (NODE ($\mathrm{L}_1$ $\bigotimes_{\mathrm{L}}$ $\mathrm{L}_2$)) = $\bigcup$  ((set $\mathrm{L}_1$) $\bigotimes$ (set $\mathrm{L}_2$))}}
\end{flushleft}

\begin{flushleft}
	\texttt{\bf{Theorem 5:}}\\
	\vspace{1pt} 
	\small{\texttt{$\vdash$ $\Omega^\vDash_{\mathrm{L}}$ ($\mathrm{L}_\mathcal{N}$::L) $\Rightarrow$ ETREE (NODE (L $\bigotimes^\mathcal{N}_{\mathrm{L}}$ $\mathrm{L}_\mathcal{N}$)) = $\bigcup$ ((set L) $\bigotimes^\mathcal{N}$ (set $\mathrm{L}_\mathcal{N}$))}}
\end{flushleft}

\noindent where the predicate function $\Omega^\vDash_{\mathrm{L}}$ covers all constraints of event outcome spaces (\textit{distinct}, \textit{disjoint}, \textit{complete} and \textit{finite}) on the given lists. 

\vspace{-3mm}
\section {ET Reduction and Partitioning Formalization}

In ET analysis \cite{papazoglou1998mathematical}, \textit{Step 2 (Reduction)} is used to model the accurate functional behavior of systems in the sense that the irrelevant branches should be removed from a complete ET of a system. To perform the reduction process, we first need to extract all possible paths from a given ET and then apply the deletion operation. For this purpose, we define the following functions in HOL4:
\begin{flushleft}
	\texttt{\bf{Definition 9:}}\\
	\vspace{1pt}
	\vspace{1pt} \small{\texttt{$\vdash$ L $\bigotimes^\mathcal{N}_{\mathrm{paths}}$ $\mathrm{L}_\mathcal{N}$ = FOLDR ($\lambda$$\mathrm{L}_1$ $\mathrm{L}_2$. $\mathrm{L}_1$ $\bigotimes_{\mathrm{paths}}$ $\mathrm{L}_2$) $\mathrm{L}_\mathcal{N}$ L}}
\end{flushleft}

\noindent where the function $\bigotimes_{\mathrm{paths}}$ takes two different lists and returns a list containing all possible ET paths in a list. To ensure consistency, we also formally verify the equivalence between Definitions~8 and 9, i.e., complete ET paths, in HOL4 as:
\begin{flushleft}
	\texttt{\bf{Lemma 1:}}\\
	\vspace{1pt} 
	\small{\texttt{$\vdash$ ETREE (NODE (L $\bigotimes^\mathcal{N}_{\mathrm{L}}$ (EVENT\_TREE\_LIST $\mathrm{L}_\mathcal{N}$))) = \\ 
	        \quad \hspace{-1.1mm} ETREE (NODE (EVENT\_TREE\_LIST (L $\bigotimes^\mathcal{N}_{\mathrm{paths}}$ $\mathrm{L}_\mathcal{N}$)))}}
\end{flushleft}
\noindent where the function \texttt{EVENT\_TREE\_LIST} is used to type-cast the normal list to \texttt{EVEN\_TREE}-typed list. \\

Now, we define a  reduction function $\boxtimes$ in HOL4 on event outcome spaces that takes a list \textit{L}, which is the output of Definition 9, a list of ET path numbers~\textit{N} to be reduced and their \textit{K}~conditional events~\textit{CE} and returns a reduced ET list~as:

\begin{flushleft}
	\texttt{\bf{Definition 10:}}\\
	\vspace{1pt} \small{\texttt{$\vdash$ L $\boxtimes$ N CE p = \\
	 \quad LUPDATE (PATH p CE) (LAST N) (DELETE\_N L (TAKE (LENGTH N-1) N))}}
\end{flushleft}

\noindent where the functions \texttt{LUPDATE}, \texttt{LAST}, and \texttt{TAKE} are the HOL4 \textit{list} theory functions to update an element, extract the last element and take a collection of elements, respectively. The function \texttt{PATH} takes a list of events from a probability space \textit{p} and extracts an intersection between the elements of the list. The function \texttt{DELETE\_N} recursively deletes \textit{N} elements from a given list corresponding to the branches that should be removed from a complete ET of a system in order to model the accurate functional behavior of systems. To ensure that the reduced ET is consistent, we formally verify the following reduction properties:\\

\noindent We first ensure that the length of ET after reduction is equal to the length of complete ET minus the number of paths that were deleted, in HOL4~as: 

\begin{flushleft}
	\texttt{\bf{Lemma 2:}}\\
	\vspace{1pt} 
	\small{\texttt{$\vdash$ (INDEX\_LT\_LEN N (L $\bigotimes^\mathcal{N}_{\mathrm{paths}}$ $\mathrm{L}_\mathcal{N}$)) $\wedge$ (LENGTH N $\geqslant$ 1) $\Rightarrow$  \\ 
	\quad \hspace{-2mm} LENGTH ((L $\bigotimes^\mathcal{N}_{\mathrm{paths}}$ $\mathrm{L}_\mathcal{N}$) $\boxtimes$ N CE p) = LENGTH (L $\bigotimes^\mathcal{N}_{\mathrm{paths}}$ $\mathrm{L}_\mathcal{N}$) - LENGTH N + 1}}
\end{flushleft}

\noindent where the function \texttt{INDEX\_LT\_LEN} ensure that each index in the given list \textit{N} is less than the length of the reduced ET list, respectively.\\

\noindent  Next, we ensure that the paths that were not reduced still exist in the reduced ET, in HOL4~as:

\begin{flushleft}
	\texttt{\bf{Lemma 3:}}\\
	\vspace{1pt} 
	\small{\texttt{$\vdash$ ($\forall$x. x $\in$ N $\Rightarrow$ i < x) $\wedge$  (SORTED ($\lambda$a b. a > b) N) $\wedge$ (LENGTH N $\geqslant$ 1) $\wedge$ \\ \quad  (INDEX\_LT\_LEN N (L $\bigotimes^\mathcal{N}_{\mathrm{paths}}$ $\mathrm{L}_\mathcal{N}$))  $\wedge$ (i $\neq$ LAST N) $\Rightarrow$  \\ 
	\quad  \hspace{-1mm} EL i ((L $\bigotimes^\mathcal{N}_{\mathrm{paths}}$ $\mathrm{L}_\mathcal{N}$) $\boxtimes$ N CE p) = EL i (L $\bigotimes^\mathcal{N}_{\mathrm{paths}}$ $\mathrm{L}_\mathcal{N}$)}}
\end{flushleft}

\noindent where the function \texttt{EL}, from the \textit{list} theory, extracts a specific element from a list. The function \texttt{SORTED} ensure that the index list \textit{N} is sorted in descending order.\\

To perform multiple reduction operations on a given ET model, we define the following recursive function, using Definition 10, in HOL4 as:

\begin{flushleft}
	\texttt{\bf{Definition 11:}}\\
	\vspace{1pt} \small{\texttt{$\vdash$ L $\boxtimes^\mathcal{N}$ (N::Ns) (CE::CEs) p = (L $\boxtimes$ N CE p) $\boxtimes^\mathcal{N}$ Ns CEs p}}
\end{flushleft}

After the ET reduction process, the next step is the partitioning of the reduced ET paths space according to the system failure and success events  \cite{papazoglou1998mathematical}. Since the output of the function $\boxtimes^\mathcal{N}$ is a list, we can define a partitioning function $\boxplus$ to extract a collection of ET paths specified in the index list \textit{N}, in HOL4 as:

\begin{flushleft}
	\texttt{\bf{Definition 12:}}\\
	\vspace{1pt} \small{\texttt{$\vdash$ N $\boxplus$ L = MAP ($\lambda$a. EL a L) N}}
\end{flushleft}

To ensure the correctness of the above function, we formally verify the following commutative property with the functions $\boxplus$ and \small{\texttt{REVERSE}}, using Definitions~11 and 12, in HOL4~as:

\begin{flushleft}
	\texttt{\bf{Lemma 4:}}\\
	\vspace{1pt} 
	\small{\texttt{$\vdash$ (REVERSE M) $\boxplus$ (L $\bigotimes^\mathcal{N}_{\mathrm{paths}}$ $\mathrm{L}_\mathcal{N}$) $\boxtimes^\mathcal{N}$ Ns CEs p) =  \\ \quad \hspace{0.4mm} REVERSE (M $\boxplus$ (L $\bigotimes^\mathcal{N}_{\mathrm{paths}}$ $\mathrm{L}_\mathcal{N}$) $\boxtimes^\mathcal{N}$ Ns CEs p)}}
\end{flushleft}

\noindent where the HOL4 function \small{\texttt{REVERSE}} returns a list in reverse order.
\vspace{-3mm}
\section {ET Probabilistic Analysis Formalization}
\label{Probability of Node and Branch}

The last step in the ET analysis \cite{papazoglou1998functional} is to determine the probability of each path occurrence in the whole ET diagram. For that purpose, we formally verify \textit{generic} probabilistic properties for \texttt{NODE}, \texttt{BRANCH}, \texttt{PATH} and $\bigotimes^\mathcal{N}_{\mathrm{L}}$ as follows: 

\begin{flushleft}
	\texttt{\bf{Theorem 6:}}\\
	\vspace{1pt} 
	\small{\texttt{$\vdash$ prob\_space p $\wedge$ 
			$\Omega^\vDash_{\mathrm{L}}$ L $\wedge$ (\small{\texttt{$\forall$y}. 
				\texttt{y $\in$ L $\Rightarrow$ y $\in$ events p}}) $\Rightarrow$ \\ \quad prob p (ETREE (NODE L)) = $\sum_{\mathcal{P}}$ p L}}
\end{flushleft}

\noindent The first assumption in the above theorem ensures that \textit{p} is a valid probability space. The next assumption is quite similar to the one described in Theorem 4. The last assumption ensures that all component states list belongs to the events space. The function $\sum_{\mathcal{P}}$ is defined to sum the probabilities of events for a given list.\\

Similarly, the probability of events in branches is the multiplication of each branch event probability with the sum of the probabilities for the next events. This can be verified in HOL4 as:

\begin{flushleft}
	\texttt{\bf{Theorem 7:}}\\
	\vspace{1pt} 
	\small{\texttt{$\vdash$ prob\_space p $\wedge$ 
			$\Omega^\vDash_{\mathrm{L}}$ L $\wedge$  
			MUTUAL\_INDEP p (X::L) $\wedge$ \\ \quad
			(\small{\texttt{$\forall$y}. \texttt{y $\in$ (X::L) $\Rightarrow$ y $\in$ events p}}) $\Rightarrow$ \\
			\quad prob p (ETREE (BRANCH X L)) = (prob p X) $\times$ $\sum_{\mathcal{P}}$  p L}}
\end{flushleft}

\noindent where the predicate function \texttt{MUTUAL\_INDEP} ensures that all events in each path of an ET are mutually independent. \\

Also,  the probability of an ET path can be verified as the multiplication of the individual probabilities of all the events associated with it, in HOL4 as:

\begin{flushleft}
	\texttt{\bf{Theorem 8:}}\\
	\vspace{1pt} 
	\small{\texttt{$\vdash$ prob\_space p $\wedge$
			MUTUAL\_INDEP p L $\wedge$ 
			(\small{\texttt{$\forall$y}. \texttt{y $\in$ L $\Rightarrow$ y $\in$ events p}}) $\Rightarrow$ \\
			\quad prob p (PATH p L)) = $\prod$ (PROB\_LIST p L)}}
\end{flushleft}

\noindent where the function $\prod$  takes a list and extracts the multiplication of the list elements while the function \small{\texttt{PROB\_LIST}} returns a list of probabilities associated with the elements of the list.\\

Additionally, we can formally verify a generic probabilistic property for the function~$\bigotimes^\mathcal{N}_{\mathrm{L}}$, in HOL4 as:

\begin{flushleft}
	\texttt{\bf{Theorem 9:}}\\
	\vspace{1pt} 
	\small{\texttt{$\vdash$ prob\_space p $\wedge$ 
			$\Omega^\vDash_{\mathrm{L}}$ ($\mathrm{L}_\mathcal{N}$::L) $\wedge$  MUTUAL\_INDEP p  ($\mathrm{L}_\mathcal{N}$::L) $\wedge$ \\ \quad (\small{\texttt{$\forall$y}. \texttt{y $\in$ ($\mathrm{L}_\mathcal{N}$::L) $\Rightarrow$ y $\in$ events p}}) $\Rightarrow$ \\  \quad
	prob p (ETREE (NODE  (L $\bigotimes^\mathcal{N}_{\mathrm{L}}$ $\mathrm{L}_\mathcal{N}$))) = $\prod$  ($\sum^{\mathrm{2D}}_{\mathcal{P}}$ p ($\mathrm{L}_\mathcal{N}$::L))}}
\end{flushleft}

\noindent where the function $\sum^{\mathrm{2D}}_{\mathcal{P}}$ is used to recursively apply the function $\sum_{\mathcal{P}}$ on a given two dimensional list. \\

Using the above theorems, we can formally verify in HOL4 that the probability of the function $\bigotimes^\mathcal{N}_{\mathrm{L}}$ is equal to 1, which returns a complete space of failure and success events:

\begin{flushleft}
	\texttt{\bf{Theorem 10:}}\\
	\vspace{1pt} 
	\small{\texttt{$\vdash$ prob\_space p $\wedge$ MUTUAL\_INDEP p ($\uparrow\downarrow$ ($L_\mathcal{N}$::L)) $\wedge$
			$\Omega^\vDash_{\mathrm{L}}$ ($\uparrow\downarrow$ ($L_\mathcal{N}$::L)) $\wedge$  \\ \quad (\small{\texttt{$\forall$y}. \texttt{y $\in$ ($\uparrow\downarrow$ ($L_\mathcal{N}$::L)) $\Rightarrow$ y $\in$ events p}}) $\Rightarrow$ \\ \quad 
	prob p (ETREE (NODE (($\uparrow\downarrow$ L) $\bigotimes^\mathcal{N}_{\mathrm{L}}$ ($\uparrow\downarrow$ [$\mathrm{L}_\mathcal{N}$])))) = 1}}
\end{flushleft}

\noindent where the function $\uparrow\downarrow$ takes a list of components and assigns both $\uparrow$ and $\downarrow$ events to each component in the given list representing operating and failing events, respectively.\\

The prime purpose of the above-mentioned formalization of ETs is to build a reasoning support for the formal analysis of reliability aspects of real-world safety-critical systems within the sound environment of HOL4. In the next section, we present the formal ET analysis of an electrical power grid and determine its reliability index to illustrate the applicability of our proposed approach.

\vspace{-3mm}
\section{Electrical Power Grid System}
\label{Sec: Case study}

An electrical power grid is an interconnected network for delivering electricity from producers to consumers. The power grid system \cite{fang2011smart} mainly consists of:~(i)~generating stations that produce electric power; (ii) electrical substations for stepping voltage up for transmission or down for distribution; (iii) high voltage transmission lines that carry power from distant sources to demand-centers; and (iv)~distribution lines that connect individual customers. With respect to the power-outage-causes study domain, the majority of the outages in the power grid are the result of events that occur on the grid transmission and distribution sides~\cite{portante2014simulation}. Therefore, a rigorous formal ET step-analysis of the power grid is essential in order to reduce the risk situation of a blackout and back-up decisions to be taken. Using our proposed ET formalization, we can model the ET for any power grid consisting of $\mathcal{N}$ transmission lines and $\mathcal{M}$ customers. Also, we can determine the System Average Interruption Frequency Index ($\mathcal{SAIFI}$), which is used by design engineers to indicate the average frequency of customers experience a sustained outage. $\mathcal{SAIFI}$ is defined as  the total number of customer interruptions over the total number of customers served~\cite{allan2013reliability}:
\begin{equation}
	\label{Eq:SAIFI}
	\centering 
	\mathcal{SAIFI}  = \frac{\sum_{{\mathcal{P}} (\mathcal{X}_{Fail}) \times \mathrm{CN}_\mathcal{X}}}{\sum_{\mathrm{CN}_\mathcal{X}}}
\end{equation}

\noindent where $\mathrm{CN}_\mathcal{X}$ is the number of customers for a certain location $\mathcal{X}$. We define a generic function $\mathcal{SAIFI}$ in HOL4 by dividing the sum of multiplying the probabilities of a collection of ET paths after reduction with the number of customers that are affected by them  over the total number of customers served~as:
\begin{flushleft}
	\texttt{\bf{Definition 13:}}\\
	\label{Definition 13}
	\vspace{1pt} \small{\texttt{$\vdash$  $\sum\nolimits_{F}$ $\mathrm{L}_\mathcal{N}$ L Ns CEs (E::Es) (CN::CNs) p = \\ 
   ($\lambda$a b. prob p (ETREE (NODE (a $\boxplus$ ((L $\bigotimes^\mathcal{N}_{\mathrm{paths}}$ $\mathrm{L}_\mathcal{N}$) $\boxtimes^\mathcal{N}$ Ns CEs)))) $\times$  b) E CN \\ +  $\sum\nolimits_{F}$ $\mathrm{L}_\mathcal{N}$ L Ns CEs Es CNs p}}
\end{flushleft}
\begin{flushleft}
	\texttt{\bf{Definition 14:}}\\
	\label{Definition 14}
  \vspace{1pt} \small{\texttt{$\vdash$ 
     $\mathcal{SAIFI}$ $\mathrm{L}_\mathcal{N}$ L Ns CEs Es CNs p = 
    ($\sum\nolimits_{F}$ $\mathrm{L}_\mathcal{N}$  L Ns CEs Es CNs p) / $\sum$ CNs}}
\end{flushleft}

\noindent where\\
\texttt{L} \hspace{2mm} : list of transmission lines (TL) modes; \quad  $L_\mathcal{N}$ : Last TL modes;  \\
\texttt{Ns} \hspace{1.5mm}: list of complete cylinders; \quad \texttt{Es} \hspace{1mm}: list of events partitioning paths; \\ 
\texttt{CEs} \hspace{-0.2mm}: list of conditional events; \quad  \texttt{CNs} : list of customer numbers\\

\begin{figure} [!b]
	\begin{center}
		\centering
		\includegraphics[scale= 0.38]{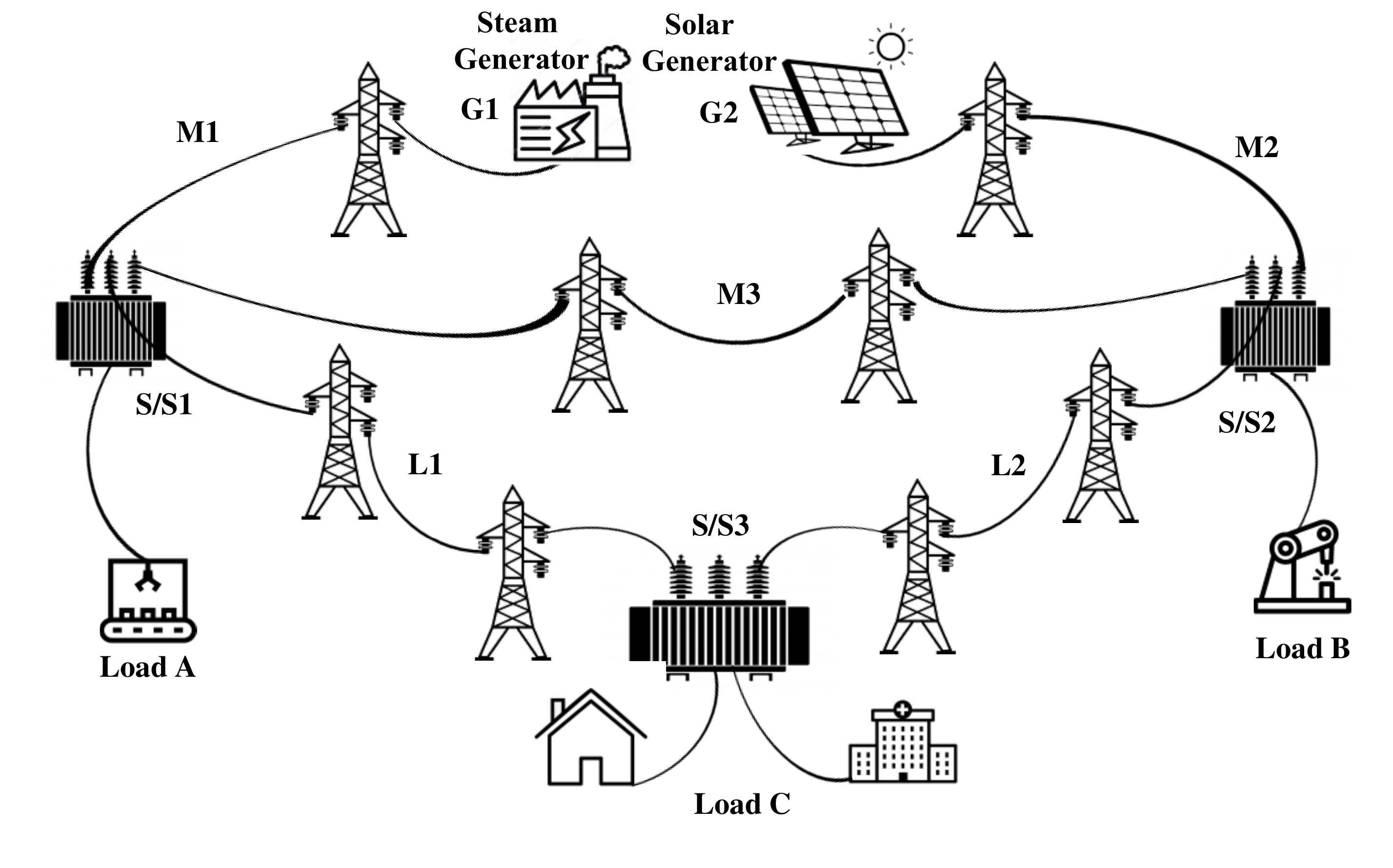} 
	\end{center}
	\caption{Electrical power grid}
	\label{CaseStudy}
\end{figure}

For instance, consider a power grid consisting of three main transmission lines~(M), two lateral transmission lines (L), two generators (G), three (two step-up and one step-down) substations (S/S) and three different loads A, B and C with the number of customers served X, Y and Z, respectively, as shown in Fig. \ref{CaseStudy}. Assume that each TL (M/L) has two operational states, i.e., operating or failing. Using our ET formalization, we can formally verify the complete ET model (32 paths) for the 5 TLs that mainly affect the reliability of the power grid, in HOL4 as: 

\begin{flushleft}
	\texttt{\bf{Lemma 5:}}\\
	\vspace{1pt} \small{\texttt{$\vdash$ ETREE (NODE 
			($\uparrow\downarrow$ [M1; M2; M3; L1]) $\bigotimes^\mathcal{N}_{\mathrm{L}}$ ($\uparrow\downarrow$ [L2]))
			=\\ \quad
			ETREE (NODE \\ \quad \hspace{0.1mm}
			[BRANCH (M1 $\uparrow$) [BRANCH (M2 $\uparrow$)
			[BRANCH (M3 $\uparrow$)\dots;
			BRANCH (M3 $\downarrow$)\dots]; \\ \qquad \quad \qquad \qquad \qquad \hspace{0.1mm} BRANCH (M2 $\downarrow$) 
			[BRANCH (M3 $\uparrow$)\dots;
			BRANCH (M3 $\downarrow$)\dots]]; \\ \quad \hspace{0.1mm} \hspace{0.1mm}
			BRANCH (M1 $\downarrow$) [BRANCH (M2 $\uparrow$) 
			[BRANCH (M3 $\uparrow$)\dots;
			BRANCH (M3 $\downarrow$)\dots]; \\ \qquad \quad \qquad \qquad \qquad \hspace{0.1mm} BRANCH (M2 $\downarrow$) 
			[BRANCH (M3 $\uparrow$)\dots;
			BRANCH (M3 $\downarrow$)\dots]]])}}
\end{flushleft}

\begin{figure} [!t]
	\begin{center}
		\centering
		\includegraphics[scale= 0.42]{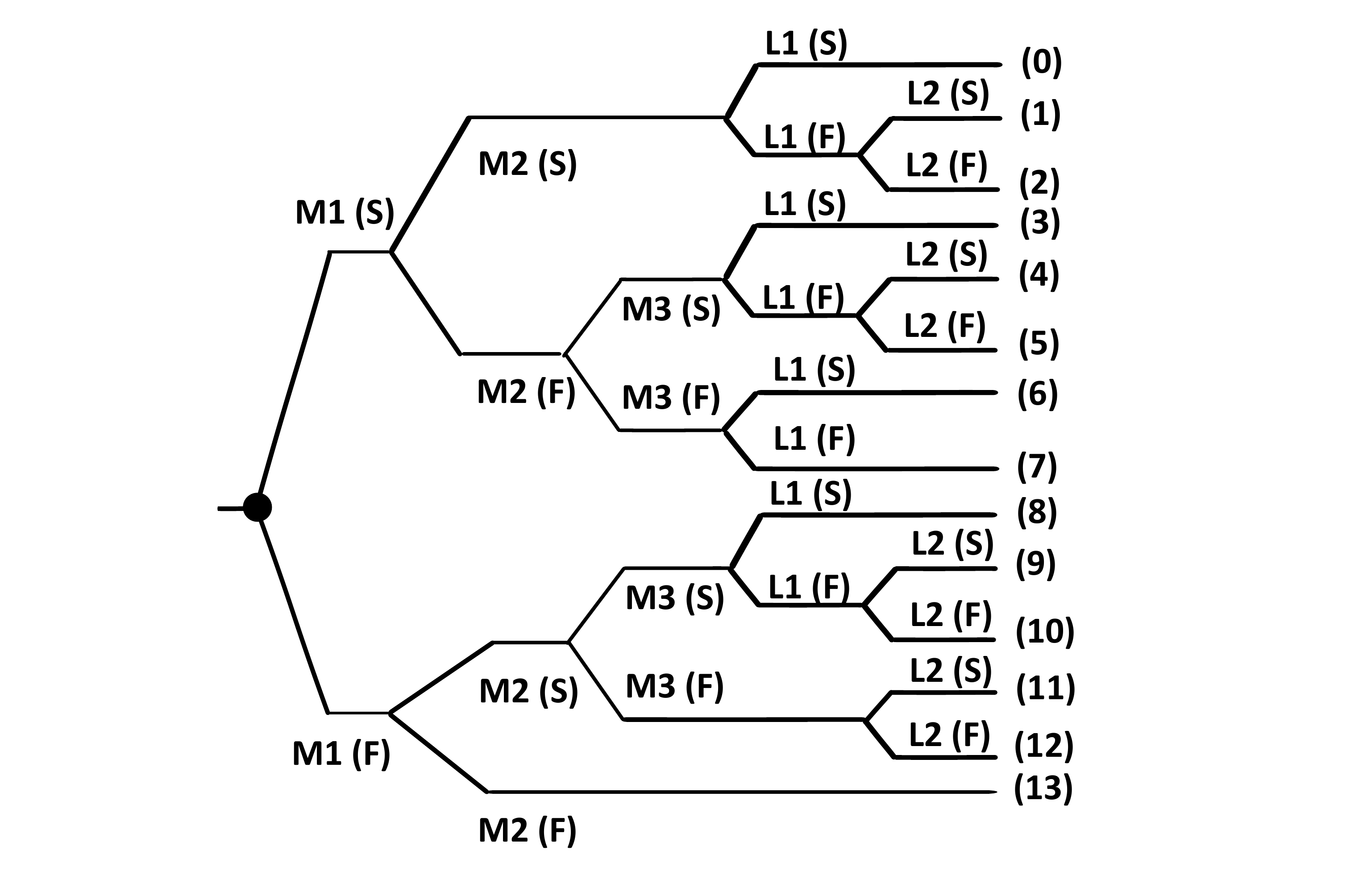} 
	\end{center}
	\caption{Reduced ET of the electrical power grid}
	\label{eventtreecasestudy}
\end{figure}

The complete ET, obtained above, can be reduced from 32 paths (0-31) to 14 paths (0-13), in the sense that the irrelevant nodes and branches are removed to model the exact logical behavior of the power grid. For instance, the paths from 31 to 24, where both M1 and M2 fail, then the likelihood of occurrence of these paths is equal to the probabilities of M1 and M2 failures only regardless of the status of other TLs. We formally verify the following reduction property to obtain the actual ET of TLs, as shown in Fig. \ref{eventtreecasestudy}, in HOL4 as: \\
\begin{flushleft}
	\texttt{\bf{Lemma 6:}}\\
	\vspace{1pt} 
	\small{\texttt{$\vdash$ ETREE (NODE (EVENT\_TREE\_LIST (($\uparrow\downarrow$ [M1; M2; M3; L1]) $\bigotimes^\mathcal{N}_{\mathrm{paths}}$ ($\uparrow\downarrow$ [L2]))  \\ \qquad  \qquad  \qquad \quad  $\boxtimes^\mathcal{N}$ [[31;30;29;28;27;26;25;24];\dots] [[M1 $\downarrow$; M2 $\downarrow$];\dots])) = \\ \quad
			ETREE (NODE \\ \qquad 
       [BRANCH (M1 $\uparrow$) \\ \qquad  \qquad  
       	   [BRANCH (M2 $\uparrow$) 
       	   [L1 $\uparrow$; BRANCH (L1 $\downarrow$) [L2 $\uparrow$; L2 $\downarrow$]];\\ \qquad  \qquad \hspace{0.1mm}
		    BRANCH (M2 $\downarrow$) 
	            [BRANCH (M3 $\uparrow$)
	            [L1 $\uparrow$; 
	            BRANCH (L1 $\downarrow$) [L2 $\uparrow$; L2 $\downarrow$]];\\ \qquad  \qquad  \qquad \qquad  \qquad \hspace{5mm}
		        BRANCH (M3 $\downarrow$) [L1 $\uparrow$; L1 $\downarrow$]]];\\ \qquad \hspace{0.1mm}
        BRANCH (M1 $\downarrow$)\\ \qquad  \qquad 
       	   [BRANCH (M2 $\uparrow$)
	           [BRANCH (M3 $\uparrow$)  
	           [L1 $\uparrow$; BRANCH (L1 $\downarrow$) [L2 $\uparrow$; L2 $\downarrow$]];\\ \qquad  \qquad  \qquad \qquad  \qquad \hspace{5mm}
	            BRANCH (M3 $\downarrow$) [L2 $\uparrow$; L2 $\downarrow$]]; M2 $\downarrow$]])}}
\end{flushleft}

Typically, we are only interested in the occurrence of certain events in ET that affect certain paths. For instance, if we consider the failure of load A, then paths 11, 12 and 13 are obtained. Similarly, a different set of paths can be obtained by observing different failures in the power grid as: (i) $\mathcal{P} (A_{Fail}) = \sum_{\mathcal{P} (paths 11,12,13)}$; (ii)~$\mathcal{P} (B_{Fail}) = \sum_{\mathcal{P} (paths 6,7,13)}$; and 
(iii) $\mathcal{P} (C_{Fail}) = \sum_{\mathcal{P} (paths 2,5,7,10,12,13)}$. \\

\noindent Therefore, the assessment of $\mathcal{SAIFI}$ can be done informally as:

\begin{equation}
	\centering 
	\mathcal{SAIFI} = \frac{\mathcal{P} (A_{Fail}) \times X + \mathcal{P} (B_{Fail}) \times  Y + \mathcal{P} (C_{Fail}) \times  Z} {X + Y + Z} 
\end{equation}

In this work, we assumed that the failure and success states of each TL is exponentially distributed. This can be formalized in HOL4 as:\\\\

\begin{flushleft}
	\texttt{\bf{Definition 15:}}\\
	\vspace{1pt} \small{\texttt{$\vdash$ EXP\_ET\_DISTRIB p X $\lambda_{X}$ =  $\forall$ t. 0 $\leq$ t $\Rightarrow$  (CDF p X t = 1 - exp (-$\lambda_{X}$ $\times$  t))}}
\end{flushleft}

\noindent where the cumulative distribution function (CDF) is defined as the probability of the event where a random variable \textit{X} has a value less or equal to a value \textit{t}, i.e.,  $\mathcal{P} (X \leq t)$. \\

Using Theorems 6-8 with the assumption that the failure and success states of each TL are exponentially distributed, we can formally verify the expression of $\mathcal{SAIFI}$ in HOL4 as follows:

\begin{flushleft}
	\texttt{\bf{Theorem 11:}}\\
	\vspace{1pt} 
	\small{\texttt{$\vdash$ $\mathcal{SAIFI}$ ($\uparrow\downarrow$ [L2]) ($\uparrow\downarrow$ [M1; M2; M3; L1]) \\ \qquad  \qquad  \quad [[31;30;29;28;27;26;25;24];\dots] [[M1 $\downarrow$; M2 $\downarrow$];\dots]\\  \qquad  \qquad  \quad
         [[11;12;13];[6;7;13];[2;5;7;10;12;13]]  [X; Y; Z] p = \\ \quad (((1 - exp (-$\lambda_{M1}$  $\times$ t)) $\times$ (exp (-$\lambda_{M2}$ $\times$ t)) $\times$ (1 - exp (-$\lambda_{M3}$ $\times$ t)) $\times$ \\ \quad \hspace{0.1mm} (exp (-$\lambda_{L2}$ $\times$ t)) +  (1 - exp (-$\lambda_{M1}$  $\times$ t)) $\times$ (exp (-$\lambda_{M2}$ $\times$ t)) $\times$ \\ \quad \hspace{0.1mm} (1 - exp (-$\lambda_{M3}$ $\times$ t)) $\times$ (1 - exp (-$\lambda_{L2}$ $\times$ t)) + \dots) $\times$ X + \\ \quad
			((exp (-$\lambda_{M1}$ $\times$ t)) $\times$ (1 - exp (-$\lambda_{M2}$ $\times$ t)) $\times$ (1 - exp (-$\lambda_{M3}$ $\times$ t)) $\times$ \\ \quad \hspace{0.1mm} (exp (-$\lambda_{L1}$ $\times$ t)) + \dots) $\times$ Y + \\ \quad
			((exp (-$\lambda_{M1}$ $\times$ t)) $\times$ (exp (-$\lambda_{M2}$ $\times$ t)) $\times$ (1 - exp (-$\lambda_{L1}$ $\times$ t)) $\times$ \\ \quad \hspace{0.1mm}  (1 - exp (-$\lambda_{L2}$ $\times$ t)) + \dots) $\times$ Z) / (X + Y + Z)}}
\end{flushleft}

To further facilitate the utilization of our proposed approach for safety engineers, we define an \textit{Auto\_$\mathcal{SAIFI}$\_ML} Standard Meta Language (SML) function that can numerically compute the above-verified expression of $\mathcal{SAIFI}$. Assume that $\lambda_{M1}$, $\lambda_{M2}$, $\lambda_{L1}$, $\lambda_{L2}$, and $\lambda_{L3}$ are 3, 2, 1, 4, 5 per year and X, Y, and Z are 250, 100, and 50 customers, respectively, then the result obtained by evaluating the $\mathcal{SAIFI}$ using \textit{Auto\_$\mathcal{SAIFI}$\_ML} is \textit{0.916173800938} interruptions/system customer. We also compared our computed result with the state-of-the-art reliability analysis tool Isograph~\cite{Isograph_tp}, which is evaluated to \textit{0.92} interruptions/system customer. It is quite evident that our proposed HOL4-based formalization approach provides the required rigor for ET analysis compared to existing simulation based approaches for system level reliability analysis. By conducting the formal ET analysis of an electrical power grid system and determining its reliability index $\mathcal{SAIFI}$, we demonstrated the practical effectiveness of the proposed ET formalization in the HOL4 theorem prover, which will help design engineers to meet the desired quality requirements. The proof-script of our proposed ET formalization and case study amounts to about 5000 lines of HOL4 code and can be downloaded from~\cite{Etree_tp}.\\
\vspace{-8mm}
\section{Conclusions}
\label{Sec: Conclusion}

In this report, we described the HOL4 formalization of ETs step-analysis using a generic \textit{list} data-type. We defined the \texttt{NODE} and \texttt{BRANCH} concepts, which can be used to model an arbitrary level of ET diagram consisting of $\mathcal{N}$ system components. We developed a formal approach to reduce ET branches, partition ET paths, and perform the probabilistic analysis based on the occurrence of certain events. For illustration purposes, we conducted the formal ET analysis of a power grid and also verified its system reliability index $\mathcal{SAIFI}$. As a future work, we plan to formalize the cascading dependencies in ETs \cite{nyvlt2012dependencies}, which will enable us to analyze hierarchical systems with many sub-system levels, based on our proposed ET formalization in the HOL4 theorem prover.  

\bibliographystyle{IEEEtran}	
\bibliography{Technical_Report}

\end{document}